\documentclass{Interspeech2024}

\usepackage{graphicx}
\usepackage{amsmath}
\usepackage{amssymb}
\usepackage{booktabs}
\usepackage{times}
\usepackage{microtype}
\usepackage{epsfig}
\usepackage[table,xcdraw]{xcolor}
\usepackage{caption}
\usepackage{float}
\usepackage{placeins}
\usepackage{color, colortbl}
\usepackage{stfloats}
\usepackage{enumitem}
\usepackage{tabularx}
\usepackage{xstring}
\usepackage{multirow}
\usepackage{xspace}
\usepackage{url}
\usepackage{subcaption}
\usepackage{xcolor}
\usepackage[hang,flushmargin]{footmisc}
\usepackage{bm}
\usepackage{comment}
\usepackage{algorithmic}
\usepackage{algorithm}
\usepackage{mathtools}
\usepackage{authblk}

\newcommand{\nbf}[1]{{\noindent \textbf{#1.}}}

\interspeechcameraready

\title{Prompt Tuning for Audio Deepfake Detection: Computationally Efficient Test-time Domain Adaptation with Limited Target Dataset}

\name[affiliation={1}]{Hideyuki}{Oiso*}
\name[affiliation={2}]{Yuto}{Matsunaga*}
\name[affiliation={2}]{Kazuya}{Kakizaki}
\name[affiliation={2}]{Taiki}{Miyagawa}

\address{
  $^1$University of Tsukuba, Japan\\
  $^2$NEC Corporation, Japan
}
\email{hideyuki@mdl.cs.tsukuba.ac.jp, yuto-matsunaga@nec.com, kazuya1210@nec.com, miyagawataik@nec.com
}


\def\thefootnote{\fnsymbol{footnote}}

\begin{document}

\maketitle
\renewcommand{\thefootnote}{\fnsymbol{footnote}}
\footnote[0]{* Equal contribution. This work was mainly completed during Oiso's internship at NEC Corporation. This paper has been accepted at Interspeech 2024.}
\renewcommand{\thefootnote}{\arabic{footnote}}

\begin{abstract}
We study test-time domain adaptation for audio deepfake detection (ADD), addressing three challenges: (i) source-target domain gaps, (ii) limited target dataset size, and (iii) high computational costs. We propose an ADD method using prompt tuning in a plug-in style. It bridges domain gaps by integrating it seamlessly with state-of-the-art transformer models and/or with other fine-tuning methods, boosting their performance on target data (challenge (i)). In addition, our method can fit small target datasets because it does not require a large number of extra parameters (challenge (ii)). This feature also contributes to computational efficiency, countering the high computational costs typically associated with large-scale pre-trained models in ADD (challenge (iii)). We conclude that prompt tuning for ADD under domain gaps presents a promising avenue for enhancing accuracy with minimal target data and negligible extra computational burden.
\end{abstract}

\section{Introduction}
Audio deepfake is a collection of deep learning techniques that create artificial speech \cite{kawa23b_interspeech}.
It can cause significant harm, including compromising the security of automatic speaker verification systems, contributing to spreading fake news, defaming an individual's reputation, and copyright violation.
Thus, developing ADD models has attracted much attention, and the ASVspoof Challenges \cite{wu15e_interspeech, delgado18_odyssey, Todisco2019ASVspoof2F, wang2020asvspoof, yamagishi2021asvspoof, liu2023asvspoof} have made significant progress.

In addressing the major challenges in ADD, we identify three primary areas of focus: (i) domain gaps in source and target data, (ii) limitation of the target dataset size, and (iii) high computational costs.

The challenge (i) arises from the domain gaps between training (source) and test (target) datasets for ADD models \cite{muller22_interspeech, tamayoflorez23_interspeech}.
They come from discrepancies in deepfake generation methods \cite{wu15e_interspeech}, recording environments (e.g., devices and surroundings) \cite{muller22_interspeech}, and languages \cite{tamayoflorez23_interspeech}, necessitating effective domain adaptation.

Specifically, we focus on \textit{test-time domain adaptation with a labeled target dataset}. 
In this context, we adapt a pre-trained model from the source domain to the labeled test dataset from the target domain, without accessing the source domain dataset during the adaptation phase. 
This approach markedly differs from prior studies in ADD under domain gaps, which focus on \textit{domain generalization}, where the target dataset is \textit{unavailable} for training \cite{shim23c_interspeech, siwen23_icassp, wen22_interspeech, davide23_icassp, zang23_interspeech}. 
In contrast, our task, emphasizing test-time domain adaptation with accessible labeled target data, remains unexplored in the context of ADD, despite its importance in practical situations. For instance, one can collect target domain data even if they are small, and one can annotate a small portion of the collected data only with a small labeling cost, which can boost the performance of ADD models in the target domain.

Regarding the challenge (ii), labeled target datasets are often small and challenging to expand due to the limited availability of additional target data. For instance, target data on new deepfake methods is hard to collect. 
In such situations, domain adaptation via full fine-tuning (i.e., updating the entire network) is ineffective, as it tends to overfit the scarce target data \cite{ma21b_interspeech}.

Regarding the challenge (iii), state-of-the-art (SOTA) ADD models require high computational costs because they often use large-scale foundation models as their feature extractors, such as wav2vec 2.0 \cite{baevski2020wav2vec} and Whisper \cite{radford2023robust}, to improve performance in test datasets \cite{kawa23b_interspeech, tak22_odyssey, wang23x_interspeech, xie23c_interspeech, rosello23_interspeech}.
This approach has been shown to be successful, but the size of foundation models is expected to be increasing in the future, leading to prohibitive computational costs for domain adaptation.

\begin{figure}[t]
    \centering
    \includegraphics[width=0.8\linewidth]{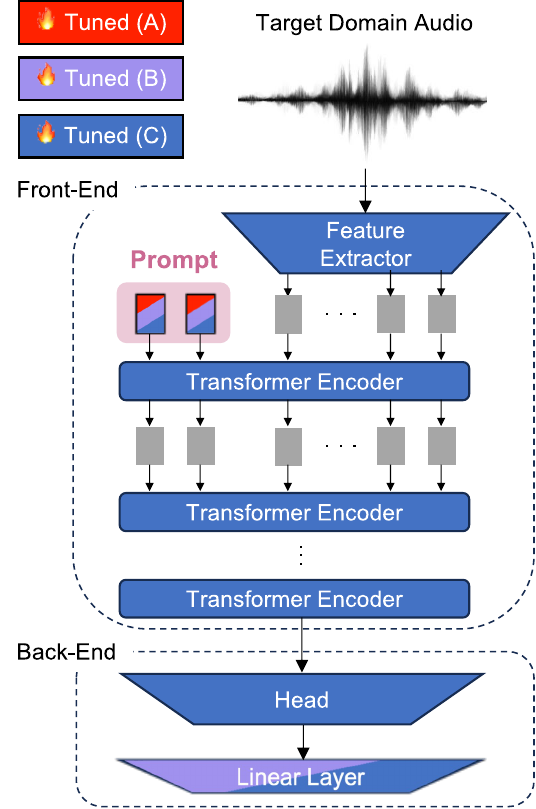}
    \caption{Proposed prompt tuning for audio deepfake detection. (A), (B), and (C) mean tuning prompt, tuning prompt \& the last linear layer, and tuning prompt \& all layers, respectively.}
    \label{fig:overview}
\end{figure}

To address these prevalent issues, we propose a plug-in style ADD method for test-time domain adaptation with a small labeled target dataset.
Our core idea is based on prompt tuning \cite{lester-etal-2021-power}, which has been used in natural language processing (NLP) and image processing and can be seen as a type of fine-tuning methods \cite{liu2023pre,jia2022visual}; however, its effectiveness for ADD is underexplored.
In prompt tuning, several trainable parameters (prompts) are inserted into the input feature (Fig.~\ref{fig:overview}) and fine-tuned on a target dataset.
Our method has the following characteristics:
(I) Our method can be easily combined with any transformer models (i.e., plug-in style), including SOTA models, as well as other fine-tuning methods, such as the linear probing (i.e., updating the last linear layer) and full fine-tuning (Fig.~\ref{fig:overview} \& Tab.~\ref{tab:results-effectiveness}).
(II) Our method can avoid overfitting small target datasets because the number of additional trainable parameters is small; in fact, our method improves the equal error rate (EER) even when the target sample size is as small as $10$ (Tab.~\ref{tab:results-sample-size}), where fully fine-tuned models immediately overfit the target dataset (Tab.~\ref{tab:results-sample-size}).
(III) The additional computational cost of our method is minimal (Tab.~\ref{tab:results-efficiency}). In our experiment, the additional trainable parameters account for only the order of  $10^{-2}\%$ or $10^{-3}\%$ compared with the number of parameters in base architectures. 
Moreover, our ablation study reveals a rapid performance saturation concerning prompt length, indicating that a prompt length of $\sim 10$ is adequate.

In summary, our contribution is threefold:
\begin{enumerate}
    \item To bridge domain gaps in ADD, we propose a plug-in-style method based on prompt tuning.
    Our method can be easily combined with any SOTA transformer models as well as other fine-tuning methods,
    reducing the EER of the target domain in many settings in our experiment (challenge (i)).
    \item Our method counters overfitting to small target datasets by utilizing a limited number of additional trainable parameters, thus overcoming dataset size limitations (challenge (ii)).
    \item Our method requires minimal extra computational resources, particularly when contrasted with full fine-tuning models, thereby addressing the issue of high computational costs due to base foundation models (challenge (iii)).
\end{enumerate}
Code for reproduction is given at \url{https://github.com/Yuto-Matsunaga/Prompt_Tuning_for_Audio_Deepfake_Detection}.
\section{Prompt tuning for audio deepfake detection}

\subsection{Problem definition: test-time domain adaptation with labeled target dataset for audio deepfake detection}
We consider test-time domain adaptation with a labeled target dataset for ADD (binary classification).
Let $\mathcal{X} \subseteq \mathbb{R}^{\Delta}$ be the audio waveform space, where $\Delta \geq 1$ is the number of sampling points. Let $\mathcal{Y} = \{Real, Fake\}$ be the binary label space.
The pre-trained ADD model, $f: \mathcal{X} \to \mathcal{Y}$ (Front-End and Back-End in Fig.~\ref{fig:overview}), parameterized by $\theta_f$, determines whether a given waveform $x \in \mathcal{X}$ is classified as $Real$ or $Fake$.
$\theta_f$ is pre-trained on a \textit{source dataset} 
$D_S=\{(x_S^i, y_S^i)\}_{i=1}^{M_S}$ 
that is sampled from a \textit{source distribution} $P_S(\mathcal{X},\mathcal{Y})$.
The \textit{target dataset} is defined as
$D_T=\{(x_T^j, y_T^j)\}_{j=1}^{M_T}$,
sampled from a \textit{target distribution} $P_T(\mathcal{X},\mathcal{Y})$. 
We assume that there exists a discrepancy between $P_S(\mathcal{X},\mathcal{Y})$ and $P_T(\mathcal{X},\mathcal{Y})$.
We aim to adapt the pre-trained model $f$ to the target distribution $P_T(\mathcal{X},\mathcal{Y})$, using post-hoc prompt tuning on $D_T$. 
This could involve introducing additional trainable parameters, or a prompt, $\theta_P$, and tuning both $\theta_P$ and $\theta_f$ on $D_T$ to minimize the empirical risk: 
$\frac{1}{M_T}\sum_{i = 1}^{M_T} \mathcal{L}(f(x_T^i  ;\theta_f, \theta_P), y_T^i)$,
where $\mathcal{L}: \mathcal{Y} \times \mathcal{Y} \to \mathbb{R}$ is an arbitrary loss function.
In our experiment, we use the class balanced loss \cite{cui2019class}, a class-imbalance-aware loss function.
This is because ADD datasets are generally imbalanced, and we want to focus on performance gains by the prompt.
Finally, it is important to note that we assume the source data $D_S$ is \textit{not} accessible at the adaptation stage due to privacy reasons, for example.
Note also that, unlike typical test-time domain adaptation, $D_T$ is labeled; i.e., $\{y_T^i\}_{i=1}^{M_T}$ as well as $\{x_T^i\}_{i=1}^{M_T}$ are available.

\subsection{Architecture} \label{sec:add_model}
The overall architecture of the ADD model used in our experiment is shown in Fig.~\ref{fig:overview}.
It is divided into two submodules: \textit{Front-End} and \textit{Back-End}, separated by dashed lines in Fig.~\ref{fig:overview}.
This Front-End-Back-End architecture is common in many SOTA models in ADD \cite{kawa23b_interspeech, tak22_odyssey, wang23x_interspeech, xie23c_interspeech, rosello23_interspeech}.
Front-End is a large-scale foundation model such as wav2vec 2.0 \cite{baevski2020wav2vec} and Whisper \cite{radford2023robust}.
It consists of the first convolutional feature extractor and the following transformer encoders.
The feature vector thus obtained is then input to Back-End.
It consists of one non-linear module (denoted by Head) and the last linear layer and performs binary classification.

\subsection{Proposed prompt tuning} \label{sec:prompt_tuning}
The prompt $\theta_P$ is inserted to the intermediate feature vectors in Front-End (Fig.~\ref{fig:overview}) during both the training phase in the target domain and the inference phase.
Specifically, a prompt is $d \times N_P$ dimensional trainable parameters, where $d$ is the dimension of a prompt, and $N_P$ is the prompt length (the number of $d$-dimensional prompt vectors).
The initialization algorithm of prompts is the same as \cite{lester-etal-2021-power}.

We compare three types of prompt tuning, denoted by Tuned (A), (B), and (C) in Fig.~\ref{fig:overview}.
During the training phase in the target domain, type (A) tunes the prompt $\theta_P$ only. 
Type (B) tunes the prompt $\theta_P$ and the last linear layer.
Type (C) tunes the prompt $\theta_P$ and all the parameters $\theta_f$ in $f$ (i.e., Front-End and Back-End).

\begin{table*}[t]
    \centering
    \caption{Statistics of target domain datasets and domain gaps from source dataset.
    \#R and \#F mean the numbers of real and fake samples in the training (Train), development (Dev), and evaluation (Eval) datasets. 
    We perform experiments with $|D_T|=10, 50, 100, $ and $1000$ (except for VCC 2020, for which $|D_T| = 1000$ is not used); i.e., we randomly sample the target training datasets $D_T$ from the original training datasets (Train).
    Each $D_T$ maintains the original real-to-fake ratio.
    In the inference phase, all data in Dev or Eval are used.
    G, E, and L in the bottom row mean the domain gaps that come from the discrepancies in deepfake generation methods (G), recording environments (E), and languages (L), respectively. 
    The source dataset $D_S$ is ASVspoof 2019 LA. 
    }
    \label{tab:dataset-statistics}
    \begin{tabular}{@{}c|c|c|c|cccc@{}}
    \toprule
        & In-The-Wild & HABLA & ASVspoof 2021 LA & \multicolumn{4}{c}{VCC 2020} \\
        &($T_1$)& ($T_2$) & ($T_3$) & English ($T_4$) & Mandarin ($T_5$) & German ($T_6$) & Finnish ($T_7$) \\
    \cmidrule(lr){2-8}
        & \#R/\#F & \#R/\#F & \#R/\#F & \#R/\#F & \#R/\#F & \#R/\#F
        & \#R/\#F \\
    \midrule
        Train & 7,935/4,776 & 11,041/27,750 & 1,676/14,788 & 40/250 & 18/194 &21/189 &23/189   \\
        Dev & 3,970/2,386 & 2,718/6,980 & 1,960/14,966 & 18/127 & 10/96 & 9/96 & 9/97     \\
        Eval & 8,058/4,654 & 9,057/23,270 & 14,816/133,360 & 42/249 & 22/190 & 20/190 &18/194  \\
    \midrule
    Gaps & G, E & G, E, L & G & G, E & G, E, L  & G, E, L & G, E, L \\
    \bottomrule
    \end{tabular}
\end{table*}
\section{Experiments}

%
\subsection{Experimental setup}
\nbf{Datasets}
We use ASVspoof 2019 LA \cite{Todisco2019ASVspoof2F} as the source dataset $D_S$ for pre-training ADD models. 
ASVspoof 2019 LA is an audio deepfake dataset from English audio samples recorded in a hemi-anechoic chamber.
For the target datasets $D_T$, we use In-The-Wild \cite{muller22_interspeech}, Hamburg Adult Bilingual LAnguage (HABLA) \cite{tamayoflorez23_interspeech}, ASVspoof 2021 LA \cite{yamagishi2021asvspoof}, and Voice Conversion Challenge (VCC) 2020 \cite{zhao2020voice}.
These four target datasets have multiple domain gaps, including deepfake generation methods (G), recording environments (E), and languages (L) (see Tab.~\ref{tab:dataset-statistics}).\footnote{
    For In-The-Wild, ASVspoof 2021 LA, and VCC 2020, which lack a predefined train-development-evaluation split, we implement random data splits, which are detailed in our code.
}
Overall, we have seven different target domains denoted by $T_1, T_2, \ldots,$ and $T_7$ in Tab.~\ref{tab:dataset-statistics}.
We perform experiments with $|D_T|=10, 50, 100, $ and $1000$ (except for VCC 2020, for which $|D_T| = 1000$ is not used); i.e., we randomly sample the target training datasets $D_T$ from the original training datasets.
Each $D_T$ maintains the original real-to-fake ratio.\footnote{The real-to-fake ratio for $|D_T| = 10$ is 1:1, not equal to the original ones, because the dataset size is too small and the ratio cannot be maintained.}
As for the development and evaluation datasets, we use the original ones.

\nbf{Pre-trained ADD models}
We use two SOTA pre-trained ADD models \cite{kawa23b_interspeech, tak22_odyssey} for the source pre-trained models (both are pre-trained on $D_S$, i.e., ASVspoof 2019 LA).
The first model \cite{tak22_odyssey}, denoted by W2V hereafter, employs wav2vec 2.0 (300M parameters) \cite{babu2021xls} and Audio Anti-Spoofing using Integrated Spectro-Temporal graph attention networks (AASIST) (297k parameters) \cite{jung2022aasist} as its Front-End and Back-End (Fig.~\ref{fig:overview}), respectively.
It is the SOTA model on ASVspoof 2021 LA, according to \cite{liu2023asvspoof}.
The second model \cite{kawa23b_interspeech}, denoted by WSP, employs Whisper (39M parameters) \cite{radford2023robust} and mel-frequency cepstral coefficients (MFCC) \cite{sahidullah2012design} as its Front-End and MesoNet (28k parameters) \cite{afchar2018mesonet} as its Back-End.
It is the SOTA model on In-The-Wild, according to \cite{kawa23b_interspeech}.
We use the official pre-trained parameters and pre-processes published by the original authors of these models.\footnote{
    \texttt{\url{https://github.com/TakHemlata/SSL_Anti-spoofing}} (W2V) and 
    \texttt{\url{https://github.com/piotrkawa/deepfake-whisper-features}} (WSP).
} 
Note that our prompt tuning can seamlessly integrate arbitrary transformer-based ADD models \cite{wang23x_interspeech, xie23c_interspeech, rosello23_interspeech}.

\nbf{Hyperparameters}
The hyperparameters are the prompt length $N_P$, learning rate $\eta$, weight decay $\lambda$, batch size $B$, and effective number $\beta$ of the class balanced loss \cite{cui2019class}.
$N_P$ is one of $1$, $5$, $10$, or $100$.
We mainly use $N_P=5$, and an ablation study is given in Tab.~\ref{tab:results-prompt-length}.
The search spaces of $\eta$, $\lambda$, $B$, and $\beta$ for W2V are $[10^{-6}$, $10^{-4}]$  (log-uniform), $[5 \times 10^{-6}$, $5 \times 10^{-4}]$ (log-uniform), $\{4, 8, 16\}$ (categorical), and $\{0.99, 0.999, 0.9999\}$ (categorical), respectively, and those for WSP are $[10^{-7}$, $10^{-5}]$  (log-uniform), $[10^{-5}$, $10^{-3}]$  (log-uniform), $\{4, 8, 16\}$ (categorical), and $\{0.99, 0.999, 0.9999\}$ (categorical), respectively.
For hyperparameter tuning, the default algorithm of Optuna \cite{akiba2019optuna} is used.
The number of training runs for hyperparameter tuning is 50 for all settings.
During these runs, the EERs evaluated on the development datasets are monitored.
The optimal hyperparameters are selected based on the best results from the 50 runs.

\nbf{Computing infrastructure and runtimes}
We use an NVIDIA Tesla V100 GPU throughout all experiments.
The training durations for W2V and WSP models on the target datasets are approximately 500 and 100 seconds, respectively, under the conditions of a prompt length of 5, batch size of 16, dataset size of 50, and 100 epochs. 
For W2V, GPU memory consumption is observed at 8.3 GBs for (A) and (B), 4.1 GBs for (B) without prompt tuning, and 17 GBs for (C) and (C) without prompt tuning.
In contrast, WSP consumes 2.3 GBs for (A) and (B), 1.7 GBs for (B) without prompt tuning, and 2.6 GBs for (C) and (C) without prompt tuning.

\subsection{Results}
We compare three fine-tuning methods illustrated in Fig.~\ref{fig:overview} ((A), (B), and (C)) with or without our prompt tuning.
Tab.~\ref{tab:results-effectiveness} presents the main result, which shows the EERs on the evaluation datasets ($N_P=5$ and $|D_T| = 50$).
Prompt tuning improves or maintains the EERs across many target domains despite its minimal additional trainable parameters (Tab.~\ref{tab:results-efficiency}); for instance, they are as small as $\sim 10^{-3}\%$ of the number of parameters in the base pre-trained model (see the upper left of Tab.~\ref{tab:results-efficiency}: (A) and W2V).
This is an advantage over full fine-tuning ((C) with or without prompt tuning), which necessitates optimizing a huge number of parameters in the base pre-trained model.
Consequently, options (A) and (B) emerge as viable alternatives under constrained computational budgets. 
Considering the recent trend towards increasing sizes of base pre-trained models, these alternative approaches could become increasingly relevant in the future.

\nbf{Ablation study: target dataset sizes}
Tab.~\ref{tab:results-sample-size} presents the ablation study results related to the size of the target dataset $|D_T|$.
It is observed that prompt tuning reduces EERs in the majority of scenarios. 
This improvement is particularly notable when the target dataset size is limited to as small as 10 samples. 
Under this condition, (A) and (B) show superior performance compared to full fine-tuning (C) both with and without prompt tuning. This superiority is attributed to the tendency of full fine-tuning to overfit the small target dataset.

\nbf{Ablation study: prompt lengths}
Tab.~\ref{tab:results-prompt-length} shows the ablation study on the prompt length $N_P$, a hyperparameter of our method.
We advise limiting the number of prompts to a small scale, such as $N_P \sim O(1)$ or $N_P \sim O(10)$, because the performance saturates rapidly.
This finding also shows that prompt tuning requires only minimal additional trainable parameters.
Finally, note that the two ablation results above are obtained from the In-The-Wild dataset ($T_1$), but the same trends are also observed for the other datasets.

\subsection{Discussion}
This paper focuses on accuracy in the target dataset rather than the source dataset. 
The latter is less relevant when considering domain gaps due to changes in recording environments and languages. 
Nevertheless, preserving source accuracy is crucial, particularly when the deepfake generation method in the target domain varies from that in the source domain; i.e., the adapted model should be robust against both the previous and new attack methods. 
To address this issue, we can easily integrate our plug-in-style method with existing techniques that maintain source accuracy while enhancing target accuracy \cite{ma21b_interspeech,zhang2023you}. 

Our proposed method assumes that the base architecture includes transformer encoders.
However, it could also be used for convolutional neural networks (CNNs) in light of recent studies of prompt tuning for CNNs \cite{bahng2022visual}.
Nevertheless, most of the SOTA models in ADD are based on transformers anyway \cite{kawa23b_interspeech, tak22_odyssey, wang23x_interspeech, xie23c_interspeech, rosello23_interspeech}; thus, our method can be applied to a wide variety of SOTA models.

\begin{table*}[t]
    \centering
    \caption{Equal Error Rates (EERs) [\%] on various target domains ($T_1,..,$ and $T_7$). Mean EERs over 12 runs with different random seeds are reported. The numbers in (...) are the standard deviations. $N_P=5$ and $|D_T|=50$ in this table. 
    PT is short for prompt tuning.
    See Fig.~\ref{fig:overview} for the definitions of (A), (B), and (C).
    Bold numbers represent better results, comparing the methods with and without PT.}
    \label{tab:results-effectiveness}
    \begin{tabular}{@{}cc|ccccccc@{}}
    \toprule
        Model & Method & $T_1$ & $T_2$ & $T_3$& $T_4$ & $T_5$ & $T_6$& $T_7$ \\
    \midrule
         \multirow{7}{*}{W2V} & (A) w/o PT& 11.4 & 7.13 & \textbf{0.84} & 16.8 & 13.1 & 4.34 & 0.00 \\
         & (A) & \textbf{10.9}(0.0) & \textbf{6.47}(0.00) & 2.53(0.00) & \textbf{7.19}(0.00) & \textbf{3.85}(0.00) & \textbf{0.789}(0.000) & 0.00(0.000)  \\
    \cmidrule{2-9}
         & (B) w/o PT & 11.4(0.0) & 6.64(0.02) & \textbf{0.823}(0.000) & 16.8(0.0) &13.1(0.0) & 4.34(0.00) & 0.00(0.00) \\
         & (B) & \textbf{10.5}(0.1) & \textbf{6.40}(0.01) & 2.53(0.00) & \textbf{9.55}(0.03) & \textbf{3.85(0.00)} &\textbf{0.789}(0.00) & 0.00(0.00) \\
    \cmidrule{2-9}
         & (C) w/o PT & 5.48(0.49) & 4.81(1.13) & \textbf{0.985}(0.021) & 3.07(2.02) & \textbf{0.00}(0.00) & 2.26(1.50) & 0.00(0.00) \\
         & (C) & \textbf{4.23}(0.47) & \textbf{2.55}(0.79) & 1.24(0.11) & \textbf{1.08}(0.45) & 0.158(0.141) & \textbf{0.632}(0.141) & 0.00(0.00)  \\
    \midrule
        \multirow{7}{*}{WSP} & (A) w/o PT & 28.7 & 20.2 & 11.7 & 0.00 & 0.00 & 0.00 & 0.00  \\
         & (A)  & \textbf{28.6}(0.0) & \textbf{19.7}(0.0) & \textbf{11.3}(0.0) & 0.00(0.00) & 0.00(0.00) & 0.00(0.00) & 0.00(0.00) \\
    \cmidrule{2-9}
         & (B) w/o PT & 28.7(0.0) & 20.2(0.0) & 11.7(0.0) & 0.00(0.00) & 0.00(0.00) & 0.00(0.00) & 0.00(0.00)  \\
         & (B) & \textbf{28.1}(0.1) & \textbf{19.6}(0.0) & \textbf{11.3}(0.0) & 0.00(0.00) & 0.00(0.00) & 0.00(0.00) & 0.00(0.00) \\
    \cmidrule{2-9}
         & (C) w/o PT & 11.8(0.3) & 17.1(0.5) & \textbf{8.46}(0.47) & 0.00(0.00) & 0.00(0.00) & 00.00(0.00) & 0.00(0.00) \\
         & (C) & \textbf{11.4}(0.3) & \textbf{16.9}(0.5) & 8.57(0.35) & 0.00(0.00) & 0.00(0.00) & 00.00(0.00) & 0.00(0.00) \\
    \bottomrule
    \end{tabular}
\end{table*}
\begin{table}[t]
    \centering
    \caption{
           Efficiency of our prompt-tuning methods.
           See Fig.~\ref{fig:overview} for the definitions of (A), (B), and (C).
           PT is short for prompt tuning.
           We use $N_P=5$ in this table. 
           ``\#Params'' means the number of trainable parameters. 
          ``Ratio [\%]'' is defined as the number of trainable parameters divided by that of the base pre-trained model (W2V or WSP).
           The ratios of additional trainable parameters are minimal. 
       } \label{tab:results-efficiency}
    {\eightpt
    \begin{tabular}{@{}c||rl|rl@{}}
    \toprule
    \multirow{2}{*}{Method} & \multicolumn{2}{c|}{W2V} & \multicolumn{2}{c}{WSP} \\
    & \multicolumn{1}{c}{\#Params} & \multicolumn{1}{c|}{Ratio [\%]} & \multicolumn{1}{c}{\#Params} & \multicolumn{1}{c}{Ratio [\%]} \\
    \midrule
          (A)  & 5,120 & 0.00161  & 1,920 & 0.0251 \\
    \cmidrule{1-5}
         (B) w/o PT & 322 & 0.000101 & 17 & 0.000274 \\
         (B) & 5,442 & 0.00171  & 1,937 & 0.0253  \\
    \cmidrule{1-5}
         (C) w/o PT & 317,837,834 & 1.00  & 7,660,881 & 1.00  \\
         (C) & 317,842,954 & 1.00161 & 7,662,801 & 1.000251  \\
    \bottomrule
    \end{tabular}
    }
\end{table}
\begin{table}[!t]
    \centering
    \caption{Ablation study on training sample size $|D_T|$ ($T_1$, W2V, and $N_P = 5$).
    Mean EERs [\%] over 12 runs with different random seeds are reported. 
    The numbers in (...) are the standard deviations.
    See Fig.~\ref{fig:overview} for the definitions of (A), (B), and (C).
    PT is short for prompt tuning.
    Bold numbers represent better results, comparing the methods with and without PT.}
    \label{tab:results-sample-size}
    \begin{tabular}{@{}c|ccc@{}}
    \toprule
        \multirow{2}{*}{Method} & \multicolumn{3}{c}{Sample size} \\
        & 10 & 100 & 1000 \\
    \midrule
    (A) w/o PT & 11.4 & 11.4 & 11.4 \\
    (A)  &\textbf{10.9}(0.0)&\textbf{11.0}(0.0)&\textbf{10.2}(0.0) \\
    \cmidrule{1-4}
         (B) w/o PT&11.4(0.0) &11.3(0.0)&10.7(0.0) \\
         (B) & \textbf{10.9}(0.0) &\textbf{10.5}(0.1)& \textbf{8.61}(0.04) \\
    \cmidrule{1-4}
         (C) w/o PT&11.4(0.8) &3.49(0.39) & \textbf{0.876}(0.158) \\
         (C)  & \textbf{11.1}(0.9)& \textbf{3.16}(0.55) & 0.999(0.124) \\
    \bottomrule
    \end{tabular}
\end{table}
\begin{table}[!t]
    \centering
    \caption{Ablation study on prompt length $N_P$ ($T_1$, W2V, $|D_T|=50$). 
    Mean EERs [\%] over 12 runs with different random seeds are reported.
    The numbers in (...) represent the standard deviations.
    See Fig.~\ref{fig:overview} for the definitions of (A), (B), and (C).
    Note that one prompt corresponds to a 1024-dimensional vector.}
    \label{tab:results-prompt-length}
    \begin{tabular}{@{}c|cccc@{}}
    \toprule
        \multirow{2}{*}{Method} & \multicolumn{4}{c}{Prompt length} \\
        & 1 & 5 & 10 & 100 \\
    \midrule
         (A)  &  12.1(0.0) & 10.9(0.0) & 10.0(0.0)&19.3(0.4) \\
         (B)  &12.1(0.0) & 10.5(0.1) & 10.2(0.0) & 22.9(0.3)\\
         (C)  &5.17(0.61)& 4.23(0.47)& 4.68(0.40) & 5.10(0.47)\\
    \bottomrule
    \end{tabular}
\end{table}

\section{Conclusion}
In our study on test-time domain adaptation for ADD, we tackle three prevalent challenges: (i) domain gaps between source and target dataset, (ii) limitation of target dataset size, and (iii) high computational costs.
To overcome these issues, we introduce a method for ADD utilizing prompt tuning in a plug-in style.
Our method can be applied to SOTA transformer models and other fine-tuning methods, to boost accuracy on target data.
Additionally, our method efficiently improves accuracy even with limited amounts of labeled target data (e.g., 10).
Furthermore, the computational cost of our method is low compared to full fine-tuning even when the base pre-trained model is huge.
Our experiments show that the proposed method improves detection performance in most cases for two SOTA models and seven domain gaps.
\clearpage
\bibliographystyle{IEEEtran}
\bibliography{99_mybib}
\end{document}